\documentclass[
preprint
]{elsarticle}

\usepackage{graphicx}
\usepackage{bm}
\usepackage{amsmath}
\usepackage{amssymb}
\usepackage{bigstrut}
\usepackage{multirow}

\newcommand{\bea}{\begin{eqnarray}}
\newcommand{\eea}{\end{eqnarray}}
\newcommand{\beq}{\begin{equation}}
\newcommand{\eeq}{\end{equation}}

\newcommand{\Eq}[1]{Eq.\,(\ref{#1})}

\begin{document}


%
%
\title{Using symmetry-adapted  optimized sum-of-products  basis functions to calculate vibrational spectra}


%
%
%
\author[metz]{Arnaud Leclerc}
\ead{Arnaud.Leclerc@univ-lorraine.fr}

\author[kingston]{Tucker Carrington \corref{corres}}
\ead{Tucker.Carrington@queensu.ca}

\address[metz]{Universit\'e de Lorraine, UMR CNRS 7565 SRSMC, \\ 1 boulevard Arago 57070 Metz, France}
\address[kingston]{Chemistry Department, Queen's University, Kingston, Ontario K7L 3N6, Canada}

\cortext[corres]{Corresponding author}


%

\begin{abstract}

Vibrational spectra can be computed without storing full-dimensional vectors by using low-rank sum-of-products (SOP) basis functions.  We introduce symmetry constraints in the SOP basis functions to make it possible to separately calculate  states  in different  symmetry subgroups.  This is done using a power method to compute eigenvalues and an alternating least squares method to optimize basis functions.  Owing to the fact that the power method favours the convergence of the lowest states, one must be careful   not to exclude basis functions of some symmetries.  Exploiting symmetry facilitates making assignments and improves the accuracy.   The method is applied to the acetonitrile  molecule.

\end{abstract}
\maketitle
%





%
%



%
%
%



\section{Introduction}

It is difficult to calculate vibrational spectra of molecules with more than four atoms without making approximations.   The most systematic and general method 
involves computing eigenvalues and eigenvectors of a basis  representation of the corresponding Hamiltonian operator. 
   The Hamiltonian  matrix is often so large that it is 
best to use iterative eigensolvers.  
  A direct product (DP) basis is convenient because it 
facilitates the evaluation of the  matrix-vector products required to use  an  iterative eigensolver
\cite{bramley1993}. 
   The basis functions are products of functions of a single
coordinate.  To use an iterative eigensolver, there is no  need to store a matrix, but it $is$ necessary to store a few vectors.    Because the size of the DP basis is
 $n^D$,  where $n$ is a representative  number of 1D basis functions for each coordinate and $D$ the number of coordinates, 
even storing vectors requires more memory than is available on most computers, if $D >$ 12 (i.e. 6 atoms).

The size of the basis, and of the vectors, can be reduced by  optimizing 1-D basis functions \cite{mctdhbook}, 
or by  forgoing the advantages of a DP basis and using instead contracted basis functions \cite{carter1988} or by  
pruning a large DP  basis  \cite{davis,avila2011_2}. 
Another strategy is to use sum-of-products (SOP) basis functions which can be represented in a
 primitive DP basis as a tensor in what is called CP format \cite{kolda2009}. 
 The key idea is that a SOP basis function can be written,
\beq    %
F(q_1, \dots, q_D) \simeq \sum_{i_1=0}^{n_1-1} \dots \sum_{i_D=0}^{n_D-1} F_{i_1 i_2\dots i_D}  
\prod_{k=1}^D \theta_{i_k}^k (q_k),
\label{wavefunction}
\eeq
where $\theta_{i_k}^k (q_k)   $    is a primitive basis function,
with
\begin{equation}
F_{i_1 i_2 \dots i_D} \simeq \sum_{\ell=1}^R  \prod_{k=1}^D f^{(\ell,k)}_{i_k} ~.
\label{sop}
\end{equation}
Storing $F_{i_1 i_2 \dots i_D}$ 
requires only storing 
one-dimensional vectors ${\bf f}^{(\ell,k)}$  \cite{rrbpm2014,phillip}.
The SOP basis functions are not contracted in the usual sense;  they are also not selected from a DP basis.    
All the primitive DP basis functions can contribute to a  single SOP basis function. 
One must choose SOP basis functions that 
span a space which includes the wavefunctions of interest.
This is the main idea of the reduced rank block power method (RRBPM) introduced  in Ref. \cite{rrbpm2014}. 
Its key advantage is that the memory cost of the method 
scales  as $\mathcal{O}(nD)$.  This makes it possible to calculate 
 energy levels and wavefunctions of molecules with 20 degrees of freedom 
 with a few GB of memory.

In  Ref.  \cite{rrbpm2014} symmetry is not used.   It is important to take advantage of symmetry for two reasons. 
 (1)   Exploiting symmetry makes it possible to assign states one computes to 
irreducible representations of the symmetry group of the Hamiltonian operator
\cite{bunkerjensen,wilson1955}. 
(2) Exploiting symmetry makes it possible to 
reduce the CPU and memory cost of computing a spectrum.     In the RRBPM case, the CPU cost is reduced because the number of power iterations decreases when  computing levels 
of a single irrep since  the effective density of states decreases.

In this letter we  introduce the basic ideas required to  exploit symmetry. 
In section~\ref{theorysec} we present the theoretical arguments and we explain how to construct efficient symmetry-constrained SOP basis functions. 
The method is then applied to a realistic 12D Hamiltonian in section~\ref{results}. 
We show that   reflection in a $\sigma_v$ plane  can be used to improve the calculation of vibrational levels of  acetonitrile, without  jeopardizing  the memory advantage of the RRBPM.

\section{Theory \label{theorysec}}

\subsection{Reduced-rank block power method (RRBPM) \label{resrrbpm} }

We assume that the potential energy surface (PES) is known and is in SOP form.  We label the coordinates  $q_k$,
 $k=1\dots D$.
There is a primitive basis,  $\theta_{i_k}^k(q_k)$, $i_k=0,\dots,n_k-1$, for each coordinate.  
The SOP basis functions are of the form of \Eq{wavefunction} with  $ F_{i_1 i_2\dots i_D} $ as in \Eq{sop}.   
The memory cost is 
 $RDn$.    Each  basis function can be written
\beq
F(q_1, \dots, q_D)=\sum_{\ell =1}^R \prod_{k=1}^D \phi^{(\ell,k)} (q_k)  
=\sum_{\ell =1}^R \prod_{k=1}^D \left( \sum_{i_k=0}^{n_k-1} f^{(\ell,k)}_{i_k} \theta_{i_k}^k(q_k) \right) ~. 
\label{Fexpansion}
\eeq
The SOP basis vectors, each of which has the form of \Eq{sop}, are made using a shifted  block power method.    This requires applying
 ${\bf{(H}}-\sigma {\bf{I}}  )$ to each vector in a block.  
$\sigma $ depends on  the block size  and the largest eigenvalue of  ${\bf{H}}$   \cite{rrbpm2014}.
Calculations are fast because only one-dimensional matrix-vector products are needed.
Each application of    
 ${\bf{(H}}-\sigma {\bf{I}}  )$ 
to a vector increases its rank by a factor of $(P+1)$, where $P$ is the number of terms in the Hamiltonian.
In the RRBPM, the rank of vectors is reduced, after each step that increases their rank,  by using  
an       alternating least square method described in Ref. \cite{beylkin2005}. 
To reduce the rank of 
  ${\bf F}^{\text{old}}$   from     $R^{\text{old}}$         to  $R^{\text{new}}$,  the ALS  algorithm  uses an iterative process to find  vectors  
 $^{\text{new}}f^{(\ell,k)}_{i_k}$  to 
 minimize    $ \parallel {\bf F}^{\text{new}} - {\bf F}^{\text{old}} \parallel $.       
This is done for each 
 coordinate  successively. This gives rise to a succession of systems of $R^{\text{new}}$   linear equations   to be solved with $n_k$ different right-hand-sides for each coordinate $q_k$.

The main steps in the algorithm are: 
\begin{itemize}
\item Apply 
 ${\bf{(H}}-\sigma {\bf{I}}  )$ 
 in parallel over a block of SOP vectors 
\item Reduce the rank using alternating least squares.
\item Every 10 to 20 iterations, orthogonalize the vectors, make a matrix representing 
 $\hat{H}$  in this  SOP  basis set.
\item Diagonalize the matrix to obtain eigenvalues and eigenvectors .    
\item Reduce the rank, update the vectors and iterate.
\end{itemize}
Unless special precautions are taken, an  SOP basis, generated as explained above,  cannot be ordered so that the corresponding Hamiltonian matrix is block diagonal. 
Our goal in this letter is to demonstrate that it is possible to make a symmetrized SOP basis and use it to obtain accurate energy levels.

\subsection{Symmetric and antisymmetric sum-of-products basis functions}

The most common way to take advantage of symmetry when computing a spectrum is to introduce symmetry-adapted basis functions in which the 
Hamiltonian matrix is block-diagonal.   
When using an iterative eigensolver this is often not the best approach.  The reason is that evaluating matrix-vector products
with a matrix obtained by representing the Hamiltonian operator in a symmetry-adapted basis may be costly.
The most straightforward approach when using an iterative eigensolver is to do different calculations for start vectors with different 
symmetries \cite{poulin1996,bramley1993}
This works because the Hamiltonian is invariant with respect to all symmetry operations and therefore 
applying the Hamiltonian to a vector does not change its symmetry. 
In this letter  we use this 
 idea  to make  symmetry-adapted    SOP basis functions from matrix-vector products.

The SOP basis functions we use transform like irreducible representations (irreps) of a sub-group of the full molecular symmetry group.    We shall explain 
the ideas for a sub-group  $\{ R, E \}$, where $E$ is the identity operation.   We assume that all of the coordinates are symmetric   or 
antisymmetric, i.e.,
\bea
& q_k \text{ symmetric   } :  & q_k \overset{R}{\rightarrow} q_k,  \\
& q_k \text{ antisymmetric  } : & q_k \overset{R}{\rightarrow} -q_k.
\eea
The primitive  1-D  basis functions are chosen so that some are even and some are odd.    %
We use a harmonic primitive basis so 
\bea
 & \theta_{i_k=even}^k(q_k) \overset{q_k \rightarrow -q_k}{\longrightarrow} \theta_{i_k=even}^k(q_k), \\
 & \theta_{i_k=odd}^k(q_k) \overset{q_k \rightarrow -q_k}{\longrightarrow} -\theta_{i_k=odd}^k(q_k).
\eea
This means that some basis functions change sign (i.e. are symmetric)  and others do not (i.e. are antisymmetric)  when $R$ is applied.   
All   $ \theta_{i_k}^k(q_k) $ of symmetric coordinates are symmetric. 
 $ \theta_{i_k=even}^k(q_k) $ of antisymmetric coordinates are symmetric. 
$ \theta_{i_k=odd}^k(q_k) $ of antisymmetric coordinates are antisymmetric. 
When $R$ is applied to a primitive D-d basis function,  $\prod_{k=1}^D \theta_{i_k}^k (q_k) $, its sign changes if the number of factors for which $i_k$ is odd is 
itself odd, otherwise applying $R$ to   $\prod_{k=1}^D \theta_{i_k}^k (q_k) $ does not change the sign.   
The  D-d basis is therefore split into two parts, one symmetric and one antisymmetric. 
%
%
%
Throughout the discussion we  use  $e$ and $o$ to indicate even (symmetric) or odd (antisymmetric)  functions (coordinates).

Knowing the symmetry properties of the primitive basis functions enables us to choose SOP start vectors that are symmetric  or antisymmetric.  
The symmetry depends  on which $\theta_{i_k}^k(q_k)$   in      Eq. \ref{Fexpansion} have nonzero  coefficients.  
An SOP vector is antisymmetric   (symmetric) if applying $R$ to the corresponding function changes (does not change) its sign.  
  This makes it possible 
to separately compute even  and odd wavefunctions.   To illustrate the ideas,   consider an example with three coordinates of which 
 $q_1$ and $q_2$ are antisymmetric  coordinates and $q_3$ is a symmetric coordinate.   
If  
we consider a single product function whose constituent   1-D  vectors
(${\bf f}^{(\ell=1,k)}={\bf f}^{(k)}$)
  are %
\beq
 f^{(1)}_{i_1} =
\left(
\begin{array}{c}
f^{(1)}_{0} \\ 
0 \\ 
f^{(1)}_{2} \\ 
0 \\ 
f^{(1)}_{4} \\ 
\vdots
\end{array}\right)
%
\;, \quad
 f^{(2)}_{i_2} = 
\left(
\begin{array}{c}
f^{(2)}_{0} \\ 
0 \\ 
f^{(2)}_{2} \\ 
0 \\  
f^{(2)}_{4} \\  
\vdots
\end{array}\right)
%
\quad
\text{ and }
\quad
%
 f^{(3)}_{i_3} = 
\left(
\begin{array}{c}
f^{(3)}_{0} \\
f^{(3)}_{1} \\ 
f^{(3)}_{2} \\ 
f^{(3)}_{3} \\ 
f^{(3)}_{4} \\ 
\vdots
\end{array}\right)
\label{ex3D}
\eeq
then the function represented by  $ \prod_{k=1}^3 
 f^{(k)}_{i_k} $ 
is necessarily symmetric because %
$ f^{(k)}_{i_k} =0,  k =1,2$ if $i_k $  is odd. 
We denote the function represented by $ \prod_{k=1}^3 
 f^{(k)}_{i_k} $  as $ee$.     
Of course,  functions made from
\beq
 f^{(1)}_{i_1} =
\left(
\begin{array}{c}
0 \\ 
f^{(1)}_{1} \\ 
0 \\ 
f^{(1)}_{3} \\ 
0 \\ 
\vdots
\end{array}\right)
\; , \quad
 f^{(2)}_{i_2} =
\left(
\begin{array}{c}
0 \\ 
f^{(2)}_{1} \\ 
0 \\ 
f^{(2)}_{3} \\  
0 \\  
\vdots
\end{array}\right)
\quad
\text{ and }
\quad
 f^{(3)}_{i_3} =
\left(
\begin{array}{c}
f^{(3)}_{0} \\ 
f^{(3)}_{1} \\ 
f^{(3)}_{2} \\ 
f^{(3)}_{3} \\ 
f^{(3)}_{4} \\ 
\vdots
\end{array}\right) ~,
\label{ex3D2}
\eeq
are also symmetric. 
We call this an $oo$ function.      
Similarly, antisymmetric  
functions can be   $eo $  or $ oe $.   
We denote these 
 $eo$ and $oe$ functions.  
  A general  
antisymmetric  
  function is a sum of  functions of both these ``types''.

If there are $M$ antisymmetric coordinates then a type will be labelled by $M$ letters and   
there are  $T=2^{M-1}$ different types. 
In formulae, types will be  labelled by an integer $t$, for example
$ee \Leftrightarrow t=1$ and $oo \Leftrightarrow t=2$. 
A general $ F_{i_1 i_2 \dots i_D}  $  is   a sum of SOP, one for each type,  
\beq
F_{i_1 i_2 \dots i_D} \simeq 
\sum_{t=1}^{T} 
\sum_{\ell_t =1}^{R_t}  
\prod_{k=1}^D f^{(\ell_t ,k)}_{i_k} ~.
\label{sopsym}
\eeq    
The number of terms of type $t$ in the  SOP is denoted $R_t$ with $\sum_{t=1}^T  R_t = R$, $R$ being the total rank of  $F_{i_1 i_2 \dots i_D}    $   .

\subsection{Symmetry-adapted  RRBPM \label{theosymSOP}}

As explained at the beginning of the previous subsection,   it should be possible to compute states of a given symmetry by using a block of start vectors each of 
which has the right  symmetry.  
For the example with three coordinates, one can calculate symmetric 
states by starting with a block of symmetric 
 vectors.   The existence of 
different types, all of which are symmetric, 
complicates this somewhat.   There are symmetric 
states  for which the largest term  in an equation like    \Eq{Fexpansion}  has a
product of an even function of $q_1$ and an even function of $q_2$ and other symmetric 
states   for which the largest term  in an equation like    \Eq{Fexpansion} has a
product of an odd  function of $q_1$ and an odd function of $q_2$.  
To favour the convergence of both types of states,  we use start vectors that are 
 low-rank SOP with one term of each type. 
Going back to the 3D example, this choice corresponds to taking a start vector with two terms, one of which is made from the vectors of 
 \eqref{ex3D}
 and the other from the vectors of    \eqref{ex3D2}.
 In this 
paper   the non-zero components are random.  
The ALS reduction is a crucial step in the RRBPM.  
The reduction begins with initial   $  f^{(\ell_t ,k)}_{i_k} ~$ values,
which are optimized to give the best $^{\text{new}}f^{(\ell_t ,k)}_{i_k} ~$.    
When random initial values are used,  
${\bf{F^{\text{new}}}}$ does not   have symmetry properties, even when  ${\bf{F^{\text{old}}}}$ does.     
It would be possible to 
obtain an  ${\bf{F^{\text{new}}}}$ of a particular symmetry   (e.g., for the 3-d example,  symmetric)
by using  initial   $  f^{(\ell_t ,k)}_{i_k} ~$ chosen so that each term in 
${\bf{F}^{\text{inital}}}$  is of one of the types consistent with that symmetry (e.g., for the 3-d example,  ee or oo).   
However, according  to the equations of Ref.  \cite{beylkin2005}, when   one attempts to reduce  a SOP of type $t'$   using  an  
${\bf{F}^{\text{inital}}}$  that is type $t$, with  $t' \ne t$,  
 one obtains  new   $  f^{(\ell_t ,k)}_{i_k} ~$  that are identically zero.     It is therefore better
to  reduce each term in 
\beq
F_{i_1 i_2 \dots i_D}   = \sum_t   F^t_{i_1 i_2 \dots i_D} = \sum_{t=1}^{T} 
\sum_{\ell_t =1}^{R_t}  
\prod_{k=1}^D f^{(\ell_t ,k)}_{i_k} 
\label{sepred}
\eeq   
with initial  $  f^{(\ell_t ,k)}_{i_k} ~$  chosen so that the initial $F^t_{i_1 i_2 \dots i_D} $ is of the same type.    
For the 3-D    example in the previous subsection  
this means that to reduce    the  $ee$ terms one must use an initial  $F^t_{i_1 i_2 \dots i_D} $  that is $ee$.

Separately reducing the terms in \Eq{sepred} also has the advantage that it enables us to ensure that  all types are represented in all vectors.
If one does not reduce separately, even  when  each   start vector is a sum of terms of different types, the  terms of the vectors generated by the RRBPM can 
 be mostly or exclusively 
vectors of one type.   This is due to  the fact that the RRBPM drives vectors towards the ground state.
To ensure that each RRBPM vector will have contributions from all types we reduce separately and  impose  ${R_t^{\text{new}}} $ values.  
This can be done by separating and then reducing and merging:
\beq
\begin{array}{c}
\sum_{t=1}^{T} \sum_{\ell_t =1}^{R_t^{\text{old}}}  \prod_{k=1}^D {}^{\text{old}}f^{(\ell_t ,k)}_{i_k} \\
\swarrow \qquad \downarrow \text{(separation) } \downarrow \qquad \searrow \bigstrut[t,b]  \\
\sum_{\ell_1 =1}^{R_1^{\text{old}}}  \prod_{k=1}^D {}^{\text{old}}f^{(\ell_1 ,k)}_{i_k}
\dots
\sum_{\ell_t =1}^{R_t^{\text{old}}}  \prod_{k=1}^D {}^{\text{old}}f^{(\ell_t ,k)}_{i_k}
\dots
\sum_{\ell_T =1}^{R_T^{\text{old}}}  \prod_{k=1}^D {}^{\text{old}}f^{(\ell_T ,k)}_{i_k} \\
\downarrow \qquad \qquad \qquad \downarrow \text{ (ALS reductions) } \downarrow \qquad \qquad \qquad \downarrow \bigstrut[t,b]   \\
\sum_{\ell_1 =1}^{R_1^{\text{new}}}  \prod_{k=1}^D {}^{\text{new}}f^{(\ell_1 ,k)}_{i_k}
\dots
\sum_{\ell_t =1}^{R_t^{\text{new}}}  \prod_{k=1}^D {}^{\text{new}}f^{(\ell_t ,k)}_{i_k}
\dots
\sum_{\ell_T =1}^{R_T^{\text{new}}}  \prod_{k=1}^D {}^{\text{new}}f^{(\ell_T ,k)}_{i_k} \\
\searrow \qquad \downarrow \text{ (merging) } \downarrow \qquad \swarrow \bigstrut[t,b] \\
\sum_{t=1}^{T} \sum_{\ell_t =1}^{R_t^{\text{new}}}  \prod_{k=1}^D {}^{\text{new}}f^{(\ell_t ,k)}_{i_k}  \\
\end{array}
\eeq

  $R^{\text{new}}  = \sum_t 
 R_t^{\text{new}} $ is fixed before the calculation is started, but how do we choose   ${R_t^{\text{new}}} $?  We have considered three strategies.   
One option  is 
\beq
R_1^{\text{new}}=R_t^{\text{new}}=\dots=R_T^{\text{new}}.    
\eeq
We  call this    reduction strategy 1. 
Reduction strategy 1 does not take into account that we should minimize errors introduced by rank reduction.   The error introduced by reduction with 
 equal partial ranks will be large for vectors dominated by one type.  
  We therefore also test strategy 2 in which 
the partial ranks $R_{t}^{\text{new}}$ 
 are adjusted,  
for each ${\bf{F}}$,   
 before each reduction so that they are proportional  to  
\beq
\vert
\langle {\bf{F}}^{\text{old},t} \vert {\bf{F}}^{\text{old}} \rangle \vert
=
\vert
\sum_{\ell_{t} =1}^{R_{t}^{\text{old}}} 
\sum_{t'=1}^{T} \sum_{\ell_{t'} =1}^{R_{t'}^{\text{old}}}  
\prod_{k=1}^D  \langle {}^{\text{old}} {\bf f}^{(\ell_{t} ,k)} 
\vert {}^{\text{old}}{\bf f}^{(\ell_{t'} ,k)}
\rangle  
\vert.
\label{weight}
\eeq
 To avoid losing one of the  types, and therefore part of the space spanned by the primitive DP basis,
 we   keep at least one term of each type even if the  weight in
 \Eq{weight} 
is very small. 
Adjusted partial ranks minimize reduction error.   However, they  tend to lock in the character of the vectors. 
Near the end of the calculation,   when the   ${\bf{F}}$   
   are nearly eigenvectors, and  changing little  when ${\bf{(H}}-\sigma {\bf{I}}  )$ is applied
there is no reason not to lock in the character.  
At the beginning of the calculation,  when the vectors change a lot after application of    ${\bf{(H}}-\sigma {\bf{I}}  )$, it is important to allow
the character of the vectors to change and minimizing reduction error is less important.     
To ensure that 
the character is not locked in too early in the calculation 
we also use strategy 3.   In strategy 3,  the partial ranks  are fixed  until the number of iterations is large enough that    
 the dominant types of the SOP  basis vectors, which are becoming closer and closer to the eigenvectors, 
and the corresponding   eigenvalues    are   stabilized  
and then we use partial ranks determined by 
\Eq{weight}.     
 Fixing the partial ranks at the beginning of the calculation ensures that some types  are not pushed out of the basis.

The memory cost of the symmetry-adapted RRBPM is very similar to that of the original RRBPM.   It 
scales as $\mathcal{O} (n D B P R^{\text{new}})$ where $n$ is a representative  number of primitive basis functions for a single  coordinate, 
$D$ is the number of coordinates, $B$ is the block size,      $P$ is the number of terms in the Hamiltonian and $R^{\text{new}}$  is the total reduction rank. 
For a subgroup with two irreps, the   symmetry-adapted RRBPM  makes it possible to reduce $B$ by about a factor of 2.  
 In other words  twice as many  eigenstates can be  obtained,  with the same amount of memory, with the symmetry-adapted RRBPM as with the original RRBPM.
Moreover, half  the components of  $f^{(\ell_{t} ,k)}_{i_k}$ for antisymmetric coordinates 
are zero and do not need to be stored in memory.

\section{Results and discussion \label{results}}


As an illustrative example we have calculated vibrational eigenstates of the acetonitrile molecule (CH$_3$CN), a 12D problem. 
We use a  Hamiltonian in normal coordinates, 
\bea
H(q_1,\dots ,q_{12})
&=& - \frac{1}{2} \sum_{i=1}^{12}  \omega_i \frac{\partial ^2}{\partial q_i^2}
+ \frac{1}{2} \sum_{i=1}^{12} \omega_i q_i^2
+\frac{1}{6} \sum_{i=1}^{12}\sum_{j=1}^{12} \sum_{k=1}^{12} \phi^{(3)}_{ijk} q_i q_j q_k \nonumber \\
&&+\frac{1}{24} \sum_{i=1}^{12} \sum_{j=1}^{12} \sum_{k=1}^{12} \sum_{\ell=1}^{12} \phi^{(4)}_{ijk\ell} q_i q_j q_k q_{\ell}.
\eea
The SOP PES is deduced from 
the quartic force field potential of B\'egu\'e \emph{et al}. \cite{begue2005}  %
 by Avila and Carrington \cite{avila2011}. 
Coordinates $q_1$ to $q_4$ are non degenerate, coordinates $q_5$ and $q_6$ are members of a doubly degenerate pair as are    
 $q_7$ and $q_8$; $q_9$ and $q_{10}$; and  $q_{11}$ and $q_{12}$. 
%


  For  CH$_{3}$CN the symmetry group is 
 $C_{3v}$ with  irreducible representations (irreps)  $A_{1}$, $A_{2}$, and $E$. 
 We use a subgroup  and divide the basis into two blocks.
The non-degenerate normal coordinates, $q_{1}$, $q_{2}$, $q_{3},$ and $q_{4}$ 
and the degenerate
 normal coordinates  $(q_{5},q_{7},q_{9},q_{11})$, are all
symmetric with respect to  reflection in  a 
$\sigma_{v}$ plane,  whereas  the degenerate normal modes  $(q_{6},q_{8},q_{10},q_{12})$, are antisymmetric with
 respect to the same   operation  \cite{henry1960}. 
The primitive basis set of products  of harmonic oscillator basis functions is split into two subsets,  one of which contains functions that 
 change sign when  $\sigma_v$ is applied and the other containing  functions that do not change sign when  $\sigma_v$ is applied.

According to section \ref{theorysec}, two separate calculations can be done. 
Symmetric  wavefunctions (which correlate with $A_1$ states and half of the $E$ states) are obtained by using 
 SOP basis functions that are  sums of terms each of which is    one of    8 possible symmetric types.  The 8 types are characterized by 
whether the  1-d functions of ($q_6,q_8,q_{10},q_{12}$) are even or odd:
\bea
(e,e,e,e); \;
(o,o,o,o); \;
(e,e,o,o); \; 
(o,o,e,e); \; \nonumber \\
(e,o,o,e); \; 
(o,e,e,o); \; 
(e,o,e,o); \; 
(o,e,o,e).
\eea
Conversely,   
antisymmetric wavefunctions (which correlate with $A_2$ states and half of the $E$ states) are obtained by using 
 SOP basis functions that are  sums of terms each of which is    one of    8 possible antisymmetric  types.  The 8 types are also characterized by 
whether their 1-d functions of ($q_6,q_8,q_{10},q_{12}$) are even or odd:
 $(e,e,e,o)$, $(o,o,o,e)$, etc.


Four calculations are presented: RRBPM without symmetry adaptation and     $R^{\text{new}} = 30  $;   
 symmetry-adapted  RRBPM with fixed partial ranks $R_t^{\text{new}}=8$ (strategy 1); symmetry-adapted
 RRBPM with  partial ranks chosen according to strategy 2 and a 
 total rank $R^{\text{new}} = 30$;
 and  symmetry-adapted     RRBPM with   partial ranks chosen according to strategy 3 and   
fixed at   $R_t=8$  at the beginning of the calculation and a total rank of  $R^{\text{new}} = 30$   
at the end of the calculation. 
In all cases the number of power iterations is 6000 and the number of ALS iterations is 10.   
  The even and odd calculations done with    symmetry-adapted RRBPM are done with a block size of 60.   A block size of 
70 is used for the calculations with the original RRBPM.
These parameters are chosen so that  
their CPU costs are nearly equal.
Separately reducing a SOP with 8 terms for each of the 8 types, i.e. a total SOP with rank 64 is a little less costly than reducing a symmetry-free SOP with rank 30 but matrix-vector products with rank 64 are more costly. 
 1000 iterations for a block size  of 60  take two days using 30     AMD Opteron(tm) 6386 SE     CPUs
at 2.8GHz. 

\begin{figure}[htp]
\centering
\includegraphics[width=0.45\linewidth]{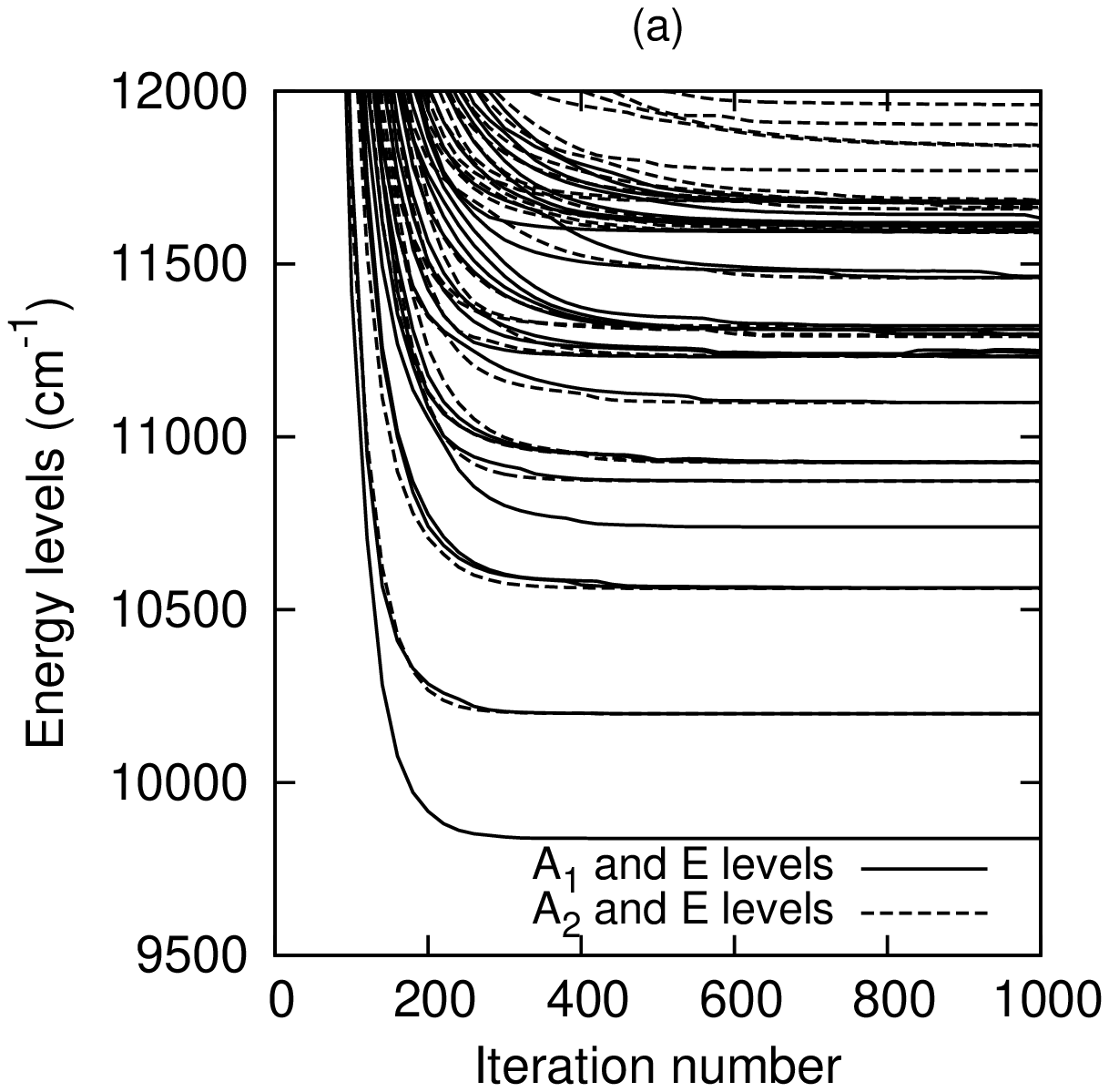}
\includegraphics[width=0.45\linewidth]{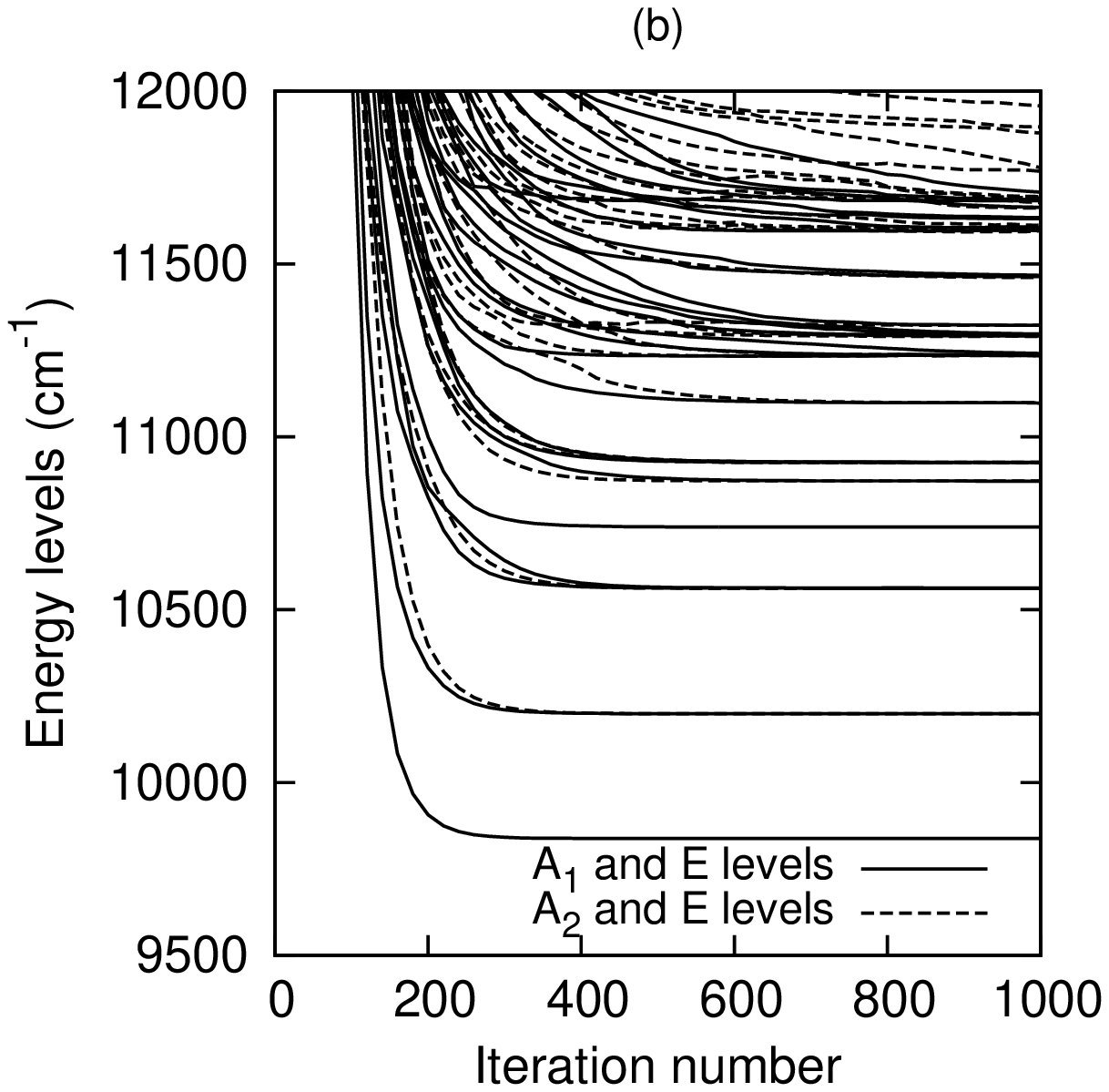}\\
\includegraphics[width=0.45\linewidth]{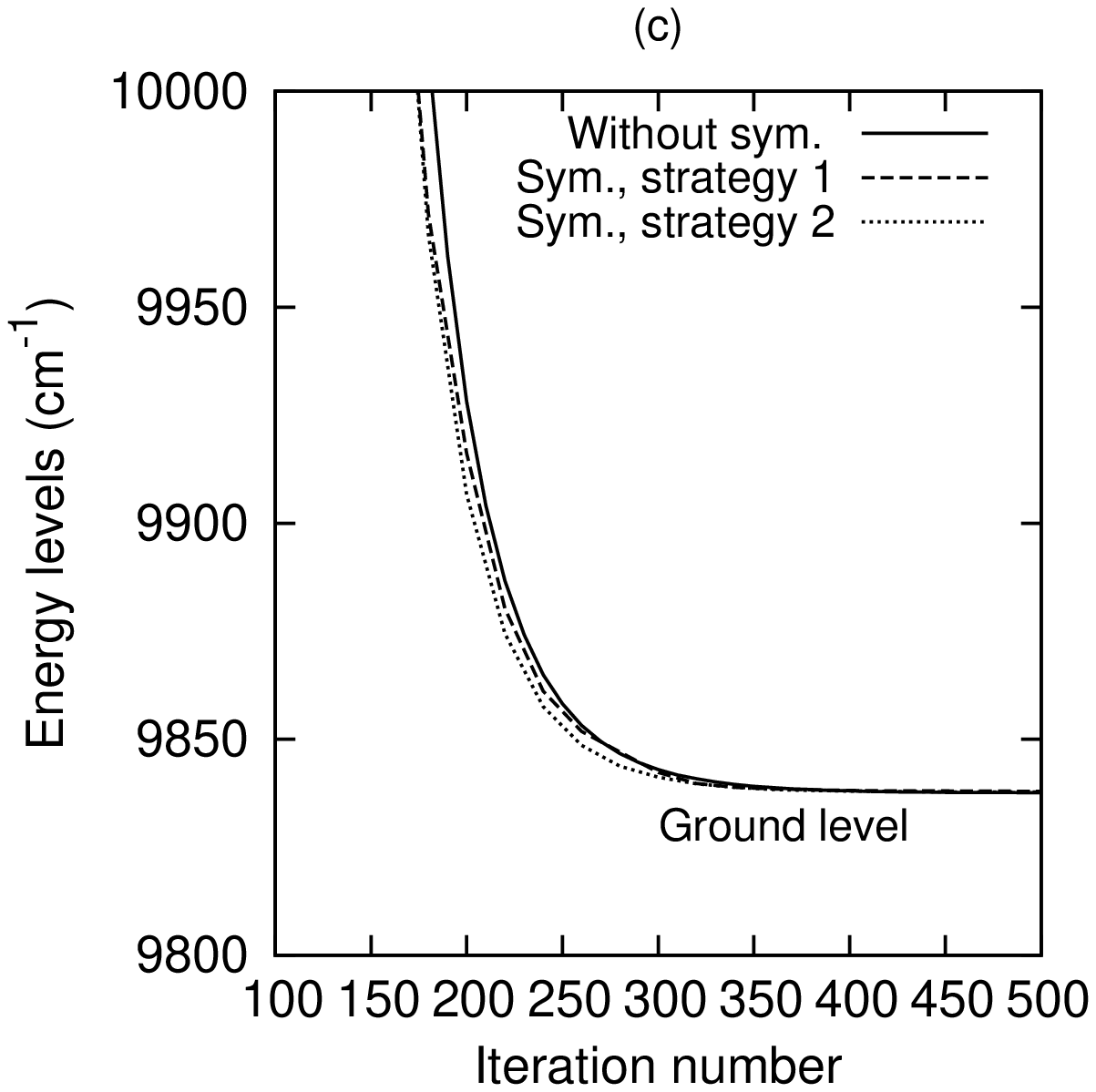}
\includegraphics[width=0.45\linewidth]{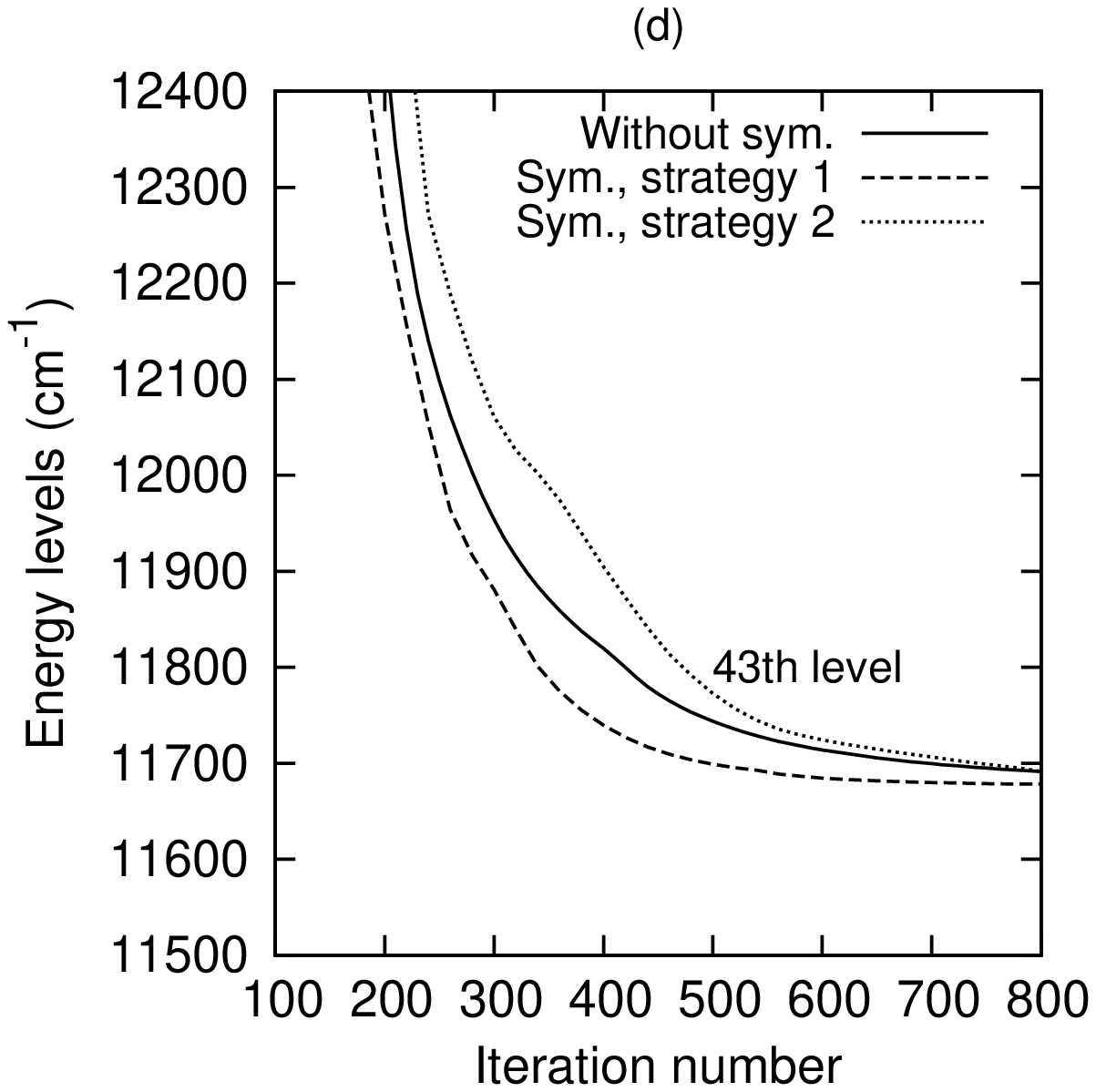}
\caption{Eigenvalues as a function of the power iteration number, using symmetry-constrained SOP with the rank reduction strategy 1 (a) or strategy 2 (b) and zooms on the first (c) and the 43th eigenvalues (d). } 
\label{convcurves}
\end{figure}


The convergence of several eigenvalues is shown  in fig. \ref{convcurves}.  They converge well with  reduction strategy 1
(fig. \ref{convcurves} (a)).   
Looking at the lowest  eigenvalues and comparing to Smolyak quadrature results of ref. \cite{avila2011},
it seems %
 that using strategy 2 improves the  accuracy. For example the %
 error 
on the first 
 transition frequency close to  $361$ cm$^{-1}$ is 0.05 cm$^{-1}$ using strategy 2 instead of 0.18 cm$^{-1}$ using strategy 1. 
However, using  strategy 2, we observe 
that some levels converge  slowly  and others are missing, even  when the number  of   power iterations is large. 
 The high-lying states of the block are particularly affected, see  Fig.  \ref{convcurves} (b).
 This appears to be due to the fact that the  partial ranks  of some types are too small.   
The 
Strategy 3 works better.  Levels computed with strategy 3 are given in table  \ref{tabwn}.

\begin{table}
\caption{Transition wavenumbers from the ZPE for CH$_3$CN in cm$^{-1}$. 
The bold values are those differing from more than 5 cm$^{-1}$ from the results of ref. \cite{avila2011}.
The braces indicate neighbouring eigenvalues whose corresponding eigenvectors are mixed in linear combinations with different symmetries. }
\hspace{-3cm}
{\small
\begin{tabular}{ccccccc}  
\hline
&&&					\multicolumn{2}{c}{RRBPM}  
					& \multicolumn{2}{c}{RRBPM}    \\
Vib.	&  			& Results 
						& \multicolumn{2}{c}{without symmetry splitting}  
											& \multicolumn{2}{c}{with symmetry splitting}    
											\\\cline{4-7} 
Assig. 	&	Sym.	& of ref. \cite{avila2011} 
						& Reduction & Reduction	& \multicolumn{2}{c}{Fixed rank for each type ($R_t=8$)} \\
		&			& (Smolyak & rank=20 		& rank=30		& \multicolumn{2}{c}{until $N_{pow}=3000$ then adaptive} \\
		& 			&	quadrature)&  Ref. \cite{rrbpm2014}	&			& \multicolumn{2}{c}{with total reduc. rank $R=30$} \\\cline{6-7} 
&&&&& subset $A_1 \oplus E$ & subset $A_2 \oplus E$ \\
\hline
ZPE 					& $A_1$	& 	-		& {9837.63} 			& {9837.51} 			& - & - \\
$\omega_{11}$ 			& $E$ 	& 360.99	& {361.18, 361.25}	& {361.07, 361.12}	& {361.06}  & {361.13} \\
$2 \omega_{11}$ 		& $E$	& 723.18 	& {723.37, 724.38}	& {723.27, 723.74}	& {723.68}  & {723.39} \\
$2 \omega_{11}$			& $A_1$ & 723.83	& {724.96}			& {724.42}			& {724.49}  & - \\
$\omega_4$				& $A_1$ & 900.66	& {900.97}			& {900.87}			& {900.90}  & - \\
$\omega_9$				& $E$ 	& 1034.13	& {1034.50, 1034.55}	& {1034.31, 1034.34}	& {1034.29} & {1034.81} \\
$3\omega_{11}$			& $A_2$ & 1086.55	& {1087.95}			& {1087.41}			& -       & {1087.25} \\
$3\omega_{11}$			& $A_1$ & 1086.55	& {1088.58}			& {1087.64}			& {1087.33} & - \\
$3\omega_{11}$			& $E$	& 1087.78	& {1090.75}, 1090.85	& {1088.81, 1088.92}	& {1088.43} & {1088.72} \\
$\omega_4+\omega_{11}$	& $E$	& 1259.88	& {1260.89, 1261.12}	& {1260.80, 1260.87}	& {1260.14} & {1260.55} \\
$\omega_3$				& $A_1$ & 1388.97	& {1391.76}			& {1391.03}			& {1390.32} & - \\
$\omega_9+\omega_{11}$	& $E$	& 1394.69	& \multirow{2}{*}{
											$
											\left.
											\begin{array}{rcl}
											{1395.74}, {1398.24} \\
											{1396.24}
											\end{array}
											\right\rbrace
											$
											}
																	& {1396.80}, {1398.51}& {1396.09} & {1395.46} \\
$\omega_9+\omega_{11}$	& $A_2$	& 1394.91	& 						& \multirow{2}{*}{
																	$
																	\left.
																	\begin{array}{rcl}
																	{\bf 1400.21},  \\
																	{\bf 1402.98}
																	\end{array}
																	\right\rbrace
																	$
																	}			&  -      & {1396.04} \\
$\omega_9+\omega_{11}$	& $A_1$ & 1397.69	& {1401.15}				& 						& {1401.03} & -       \\
$4\omega_{11}$			& $E$	& 1451.10	& {1452.92}, {\bf1458.62}& {1452.09, 1452.26}	& {1452.40} & {1452.02} \\
$4\omega_{11}$			& $E$	& 1452.83	& \multirow{2}{*}{
											$
											\left.
											\begin{array}{rcl}
											{1456.24}, {\bf1460.80} \\
											{\bf 1464.40}
											\end{array}
											\right\rbrace
											$
											}						& {1454.22, 1454.96}	& {1454.61} & {1453.81} \\	
$4\omega_{11}$			& $A_1$	& 1453.40	& 						& {1455.37}			& {1455.46} & -       \\
$\omega_7$				& $E$	& 1483.23	& {1483.52, 1483.51}	& {1483.43, 1483.47}	& {1483.46} & {1484.22} \\
$\omega_4+2\omega_{11}$	& $E$	& 1620.22	& {1621.34}, 1623.05	& {1620.98, 1622.06}	& {1621.79} & {1620.90} \\
\vdots & \vdots & \vdots & \vdots & \vdots & \vdots &\vdots  \\ 
$\omega_9+2\omega_{11}$ & $E$ & 1759.772 &  {\bf 1780.66, 1780.86} & {\bf 1771.64, 1781.84} & {\bf 1771.25} & {\bf 1767.08} \\
\vdots & \vdots & \vdots & \vdots & \vdots & \vdots &\vdots  \\
$5\omega_{11}$			& $E$ 	& 1816.80	& {\bf 1823.34, 1830.31}	& {1818.29, 1818.57}	& {1818.42} & {1818.72} \\
$5\omega_{11}$			& $A_2$ & 1818.95	& {\bf 1827.34}			& {1823.01}			& -       & {1820.93} \\
$5\omega_{11}$			& $A_1$ & 1818.95	& {\bf 1832.19}			& {\bf 1823.98}			& {1821.06}  & -         \\
$5\omega_{11}$			& $E$	& 1820.03	&  1823.87, {\bf 1828.40}	& {1821.91}, {1823.55} & {1822.58} & {1822.27} \\   
$\omega_7+\omega_{11}$	& $A_2$ & 1844.23	& \multirow{2}{*}{
											$
											\left.
											\begin{array}{rcl}
											{1845.57} \\
											{1846.85}, {\bf 1849.44}
											\end{array}
											\right\rbrace
											$
											}					& \multirow{3}{*}{
																$
																\left.
																\begin{array}{rcl}
																{1845.89}, \\
																{1847.45, 1848.12} \\
																{\bf 1850.66}
																\end{array}
																\right\rbrace
																$
																}				& -       & {1845.39} \\
$\omega_7+\omega_{11}$	& $E$	& 1844.33	& 					&					& {1845.95} & {1847.24} \\
$\omega_7+\omega_{11}$	& $A_1$ & 1844.69	& {1848.14}		&					& {1848.07}& -       \\
\vdots &\vdots & \vdots & \vdots & \vdots & \vdots &\vdots \\
\hline
\end{tabular}
}
\label{tabwn}
\end{table}

The most important advantage of the symmetry-adapted RRBPM is that it provides symmetry labels.    It also improves the accuracy of some of the 
levels reported in table \ref{tabwn}. 
Some wavefunctions obtained with  the non-symmetrized RRBPM calculation  are nearly linear combinations of a few of  the 
essentially exact wavefunctions of \cite{avila2011} 
with similar energies but different symmetries     \cite{rrbpm2014}.   
 These poorly converged  states are enclosed in  braces in the table. 
The symmetry-adapted RRBPM calculation does a better job on these states for several reasons.   1) Owing to the    
 symmetry-adapted  SOP basis, there is no mixing between 
symmetric and antisymmetric states.  
     2) In the symmetry-adapted case the effective 
density of states is lower and therefore fewer power iterations are required to achieve a given convergence error. 

\begin{figure}[htp]
\centering
\includegraphics[width=\linewidth]{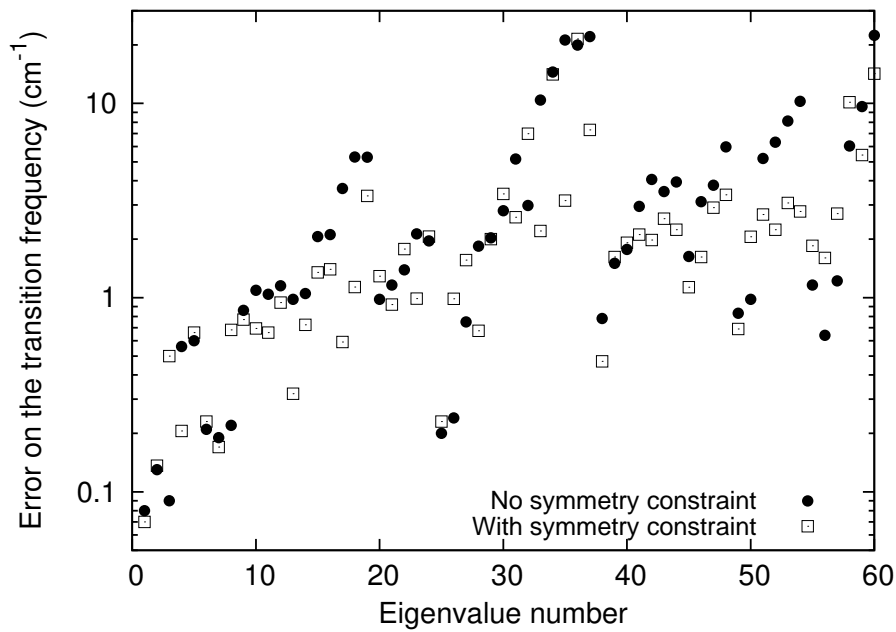}
\caption{Differences between transition wavenumbers obtained using the RRBPM and the "exact" results of ref. \cite{avila2011}, with or without symmetry constraints.} 
\label{error}
\end{figure}

Differences between the levels  obtained with Smolyak quadrature \cite{avila2011}  and levels obtained with the RRBPM, with and without symmetry adaptation, are 
 reported in fig. \ref{error}. 
For most eigenvalues, both    differences   are less than  5 cm$^{-1}$.   
Some errors are larger.   
The errors could be reduced by using contracted basis functions \cite{phillip},   
larger ranks, and more power iterations. 
The accuracy of the first several dozen 
 eigenvalues is similar  with and without symmetry adaptation.
Some of the higher levels are more accurate with  symmetry adaptation.   The non symmetry-adapted eigenvectors corresponding to 
these levels are nearly linear combinations of exact eigenvectors of different symmetries. 
 The symmetry-adapted method always has the advantage that it allows one to assign levels.

\section{Conclusion}

In this letter we introduce a symmetry-adapted version of the RRBPM.   The memory cost of the  symmetry-adapted version  is similar to that of the    original RRBPM.
It is clear that by using a block of starting vectors of a given symmetry 
 it is possible to compute states of that symmetry.  This enables one to obtain 
states with symmetry labels and accelerates convergence of the power method.  
Unless one is careful, it     is possible to make SOP basis vectors in which some ``types'' are missing or underrepresented.   We have developed several
strategies for dealing with this problem and shown that they are effective. 
  Starting vectors of a given symmetry are easily made from  
 $f^{(\ell_{t} ,k)}_{i_k}$ that have appropriate symmetries.   
Accuracy can be improved by using contractions, increasing the ranks, and increasing the number of power iterations.     Better eigensolvers  \cite{davidson1975,ribeiro2005} 
might also be adapted to the SOP format.   
The key advantage of the RRBPM is its low memory cost:  less than 1GB is required  for a 12D problem.     The memory cost is actually reduced by introducing symmetry adaptation, but 
the memory cost is so low that the reduction  is unimportant.    The major advantage of the symmetry-adapted approach is that levels are obtained with symmetry labels.   Accuracy is 
also somewhat improved.  Ideas similar to those of this letter could be used with any subgroup for which it is possible to make  ${\bf{F}}$ 
that transform like irreps.    It should be possible to use projection operators to generate states of all symmetries from one set of 
matrix-vector products, as in    the symmetry adapted Lanczos algorithm  \cite{wang2001,chen2001},  
 this will obviate the need to use symmetry adapted coordinates, which we exploit in this Letter.   

\section*{Acknowledgments}
Some  of the calculations were done   on computers purchased with a grant for the Canada Foundation for Innovation.  This research was funded by the 
Natural Sciences and Engineering Research Council of Canada.

\bibliographystyle{elsarticle-num}
\bibliography{symmetric_RRBPM}

\end{document}